\documentclass[11pt,a4paper]{article}

\usepackage{graphicx}
\usepackage{hyperref}
\usepackage{booktabs}
\usepackage{longtable}
\usepackage{amsmath}
\usepackage{amssymb}
\usepackage{float}
\usepackage{natbib}
\usepackage{url}
\usepackage[utf8]{inputenc}

\usepackage[margin=1in]{geometry}

\hypersetup{
    colorlinks=true,
    linkcolor=blue,
    citecolor=blue,
    urlcolor=blue,
    hypertexnames=false
}

\title{ThS-QCA: Threshold-Sweep Qualitative Comparative Analysis in R}

\author{
Yuki Toyoda\\
Hosei University\\
Tokyo, Japan\\
\texttt{yuki.toyoda.ds@hosei.ac.jp}\\
ORCiD: \href{https://orcid.org/0009-0007-4477-1383}{0009-0007-4477-1383}
}

\date{February 2026}

\begin{document}

\maketitle

\begin{abstract}
Qualitative Comparative Analysis (QCA) requires researchers to choose calibration and dichotomization thresholds, and these choices can substantially affect truth tables, minimization, and resulting solution formulas. Despite this dependency, threshold sensitivity is often examined only in an ad hoc manner because repeated analyses are time-intensive and error-prone. We present ThS-QCA (Threshold-Sweep QCA), a methodological framework for treating calibration thresholds as explicit analytical variables in Qualitative Comparative Analysis, implemented in the \texttt{ThSQCA} R package. ThS-QCA automates threshold-sweep analyses through four sweep functions (otSweep, ctSweepS, ctSweepM, dtSweep) to explore outcome thresholds, single-condition thresholds, multi-condition threshold grids, and joint outcome--condition threshold spaces, respectively. The package integrates with the established CRAN package QCA for truth table construction and Boolean minimization, while returning structured S3 objects with consistent print/summary methods and optional detailed results. The package also supports automated Markdown report generation and configuration-chart output to facilitate reproducible documentation of cross-threshold results.
\end{abstract}


\tableofcontents

\section{Introduction}\label{introduction}

\subsection{The threshold dependency problem in QCA}\label{the-threshold-dependency-problem-in-qca}

Qualitative Comparative Analysis (QCA) has become a standard method for analyzing complex causal relationships in social science research \citep{ragin1987, ragin2008, schneider2012}. QCA can handle equifinality and conjunctural causation by identifying combinations of conditions that are sufficient (or necessary) for an outcome.

A fundamental and critical step in QCA is calibrating raw data into set membership scores. In crisp-set QCA (csQCA), this process involves dichotomizing continuous variables using thresholds. Even in fuzzy-set QCA (fsQCA), calibration anchors (full membership, crossover point, and full non-membership) similarly require threshold decisions \citep{schneider2012}.

Despite the crucial role of thresholds, systematic examination of their impact remains limited. Researchers typically report results for a single threshold setting, and robustness checks---when conducted---are often relegated to appendices. This practice obscures a critical question: How stable are QCA findings to reasonable threshold variations? Previous research has demonstrated that QCA results can be highly sensitive to parameter choices, including calibration thresholds \citep{krogslund2015, thiem2016}.

This issue is not merely technical. In applied research, thresholds often reflect substantive decisions about what constitutes ``high'' or ``low'' values. Whether the threshold for ``high customer satisfaction'' is set at 7 or 8 on a satisfaction scale carries different implications. ThS-QCA\footnote{The methodology is termed ThS-QCA (Threshold-Sweep QCA) in the companion methodological paper. The R package implementing this framework was previously distributed on CRAN as \texttt{TSQCA} and has been renamed to \texttt{ThSQCA} (v2.0.0) to ensure naming consistency.} enables researchers to examine how such decisions affect their conclusions.

\subsection{Existing approaches and their limitations}\label{existing-approaches-and-their-limitations}

Current QCA software provides limited functionality for threshold sensitivity analysis. Table~\ref{tab:existing-packages} presents a comparison of major packages.

\begin{table}[ht]
\centering
\small
\caption{Threshold sensitivity features in existing packages}
\label{tab:existing-packages}
\begin{tabular}{p{2.5cm}p{5cm}p{4cm}}
\toprule
Package & Threshold Sensitivity Feature & Limitation \\
\midrule
\texttt{QCA} \citep{dusa2019} & Manual recalibration required & Systematic exploration difficult \\
\texttt{SetMethods} \citep{oana2018} & Sensitivity range for calibration anchors & Binary thresholds not covered \\
fs/QCA \citep{ragin2016} & No feature & --- \\
Tosmana \citep{cronqvist2019} & No feature & --- \\
\bottomrule
\end{tabular}
\end{table}

The \texttt{SetMethods} package provides functionality to assess sensitivity ranges for consistency cutoffs and calibration anchors following the robustness test protocol proposed by Oana and Schneider. These tools primarily focus on evaluating local robustness around a given calibration or cutoff choice. However, no existing package offers a systematic sweep-based exploration of dichotomization thresholds in csQCA across a predefined threshold space. As a result, researchers typically rely on manually repeating analyses for different threshold settings, a process that is both time-consuming and prone to error.

\subsection{Contributions of ThS-QCA}\label{contributions-of-tsqca}

The \texttt{ThSQCA} package addresses this methodological gap in six ways:

\begin{enumerate}
\item \textbf{Automation of threshold sweeps}: A single function call explores the entire threshold range specified by the user. This enables efficient execution of analyses across a large, predefined grid of threshold settings.

\item \textbf{Structured output}: Results are organized as S3 objects with \texttt{print()} and \texttt{summary()} methods, facilitating cross-threshold comparisons. Summary tables, detailed solution objects, and reproducibility settings are included.

\item \textbf{Integration with the QCA package}: ThS-QCA uses the core functions of the established \texttt{QCA} package \citep{dusa2019} internally, including support for negated outcomes (\texttt{outcome = "\textasciitilde{}Y"}). This ensures methodological consistency and allows researchers to benefit from updates to the \texttt{QCA} package.

\item \textbf{Report generation}: Structured markdown reports are automatically generated from analysis results, improving reproducibility and ease of reporting.

\item \textbf{Configuration table generation}: Fiss-style configuration tables \citep{fiss2011} with standard QCA notation ($\bullet$ for presence, $\otimes$ for absence) can be generated automatically, supporting Unicode, ASCII, and LaTeX output formats for direct use in academic publications.

\item \textbf{Multiple solution handling}: For multiple equivalent solutions, the package identifies Essential Prime Implicants (EPIs) and Selective Prime Implicants (SPIs).

\end{enumerate}

The package transforms thresholds from fixed parameters to analytical variables, enabling researchers to systematically map the solution space. This shift in perspective---from treating thresholds as ``assumptions'' to treating them as ``objects of study''---opens additional analytical possibilities.

\subsection{Paper organization}\label{paper-organization}

The remainder of this paper is organized as follows. Section~\ref{theoretical-background} reviews the set-theoretic foundations of QCA and clarifies why calibration thresholds constitute a critical source of analytical dependency. Section~\ref{package-design} describes the design philosophy of the ThS-QCA package and its relationship to the established \texttt{QCA} package. Section~\ref{main-functions} introduces the main functions of ThS-QCA and demonstrates their basic usage through fully reproducible examples based on the bundled sample dataset. Section~\ref{demonstration-of-advanced-output-and-reporting-capabilities} presents the advanced output structures and reporting capabilities of ThS-QCA, illustrating how results from threshold-sweep analyses can be systematically summarized, visualized, and exported for reproducible reporting. Section~\ref{comparison-with-existing-packages} compares ThS-QCA with existing packages. Section~\ref{discussion} discusses scope, limitations, and potential extensions. Section~\ref{conclusion} concludes the paper.

\section{Theoretical background}\label{theoretical-background}

\subsection{Calibration and threshold dependency}\label{calibration-and-threshold-dependency}

This section does not aim to provide a comprehensive introduction to QCA. Instead, it focuses exclusively on the set-theoretic concepts necessary to understand threshold sweep analysis as implemented in ThS-QCA. Readers seeking a full theoretical treatment of QCA are referred to established textbooks and reviews \citep{schneider2012, oana2021book, thomann2020}.

QCA examines subset relationships between condition configurations and outcomes, quantified through consistency and coverage indices \citep{ragin2006, schneider2012}. For a comprehensive treatment of set-theoretic foundations and their application in QCA, see \citet{schneider2012} and \citet{dusa2019}.

A critical step in crisp-set QCA is dichotomizing continuous variables using thresholds:

\[X_{binary} = \begin{cases} 1 & \text{if } X \geq \tau \\ 0 & \text{otherwise} \end{cases}\]

The choice of threshold $\tau$ affects all stages of QCA analysis:

\textbf{Stage 1: Set membership determination}

Changing threshold $\tau$ directly changes each case's set membership. For example, changing from $\tau=7$ to $\tau=8$ reclassifies cases with $X=7$ from ``high X'' to ``low X.''

\textbf{Stage 2: Truth table construction}

Changes in set membership alter the number of cases (n) and the consistency score (incl) for each configuration in the truth table. Specifically:

\begin{itemize}
\item Configurations falling below the frequency threshold (n.cut) are excluded from analysis
\item Configurations falling below the consistency threshold (incl.cut) are treated as outcome=0
\end{itemize}

\textbf{Stage 3: Boolean minimization}

Changes to the truth table yield different prime implicants through the Quine-McCluskey algorithm or enhanced methods such as Consistency Cubes \citep{dusa2018}, altering the final solution expression.

\subsection{The concept of hierarchical sufficiency}\label{the-concept-of-hierarchical-sufficiency}

When varying outcome thresholds, we often observe a pattern we call ``hierarchical sufficiency'': as the required outcome level increases, the configurations sufficient for that outcome become more complex, while their number decreases.

\begin{figure}[H]
\centering
\includegraphics[width=0.95\linewidth]{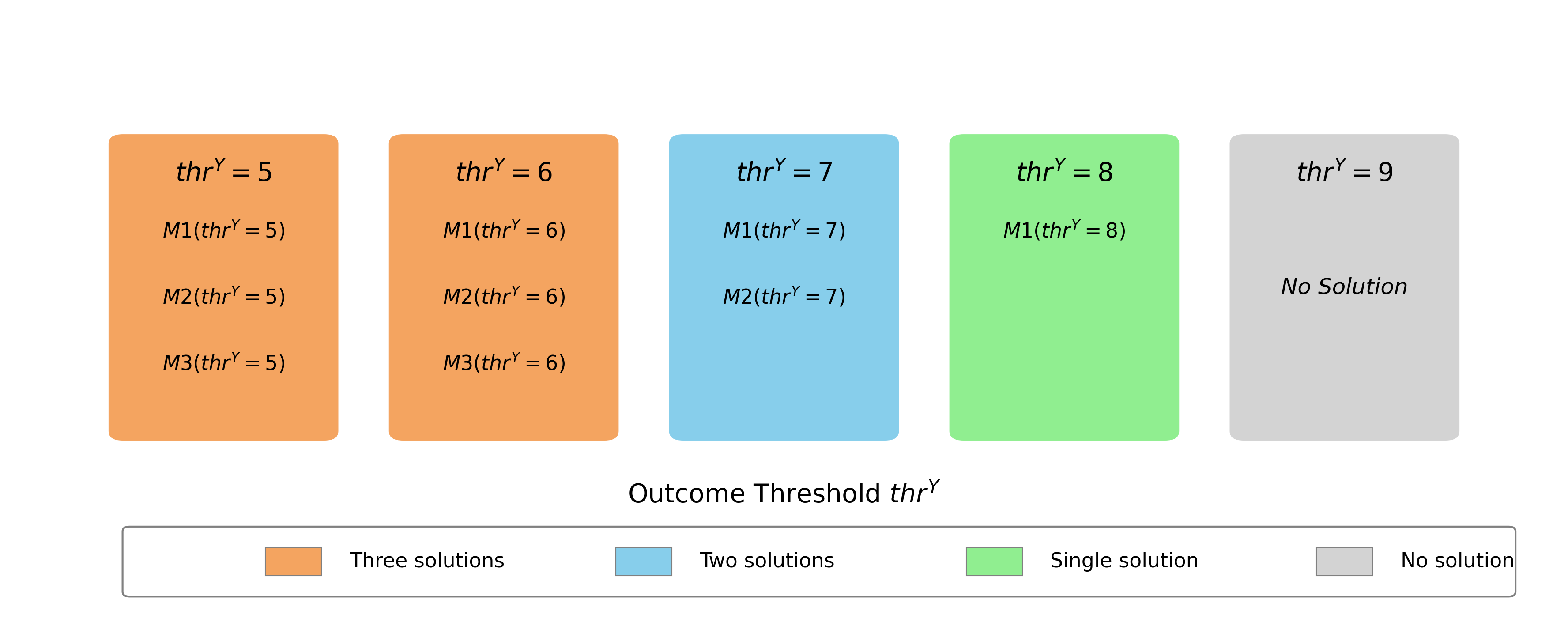}
\caption{Conceptual illustration of hierarchical sufficiency across outcome thresholds. As the outcome threshold increases from thrY = 5 to thrY = 9, the number of sufficient paths decreases from three (M1, M2, M3) to none. This figure is schematic and intended to illustrate the general logic of hierarchical sufficiency; actual empirical results vary by dataset.}
\label{fig:fig1}
\end{figure}

Figure~\ref{fig:fig1} illustrates this concept schematically. At lower thresholds (thrY = 5), multiple relatively simple configurations may suffice---indicated by multiple paths (M1, M2, M3). As thresholds increase, paths begin to disappear, and remaining paths require more conditions. At very high thresholds (thrY = 9), no configuration may meet the consistency requirements.

This pattern aligns with the intuitive understanding that achieving higher outcome levels requires more demanding combinations of conditions. Hierarchical sufficiency is also practically important, because it yields strategic implications according to target levels. For moderate targets, simple interventions may suffice, whereas ambitious targets typically require more comprehensive approaches.

We introduce this concept as a descriptive pattern observed through threshold sweep analysis, not as a formal theoretical law. The term does not imply a nested-set relationship in the strict set-theoretic sense, but rather describes an empirical pattern observed across threshold levels.

\subsection{Relationship to existing robustness research}\label{relationship-to-existing-robustness-research}

Discussion of robustness in QCA has been developed most prominently by \citet{skaaning2011} and \citet{oana2021}. Related work on sensitivity diagnostics has sought to formalize how QCA results respond to parameter perturbations \citep{thiem2016}. More recently, \citet{baumgartner2022} investigated the concept of robust sufficiency and, through simulation experiments, benchmarked the performance of different QCA solution types in recovering robustly sufficient conditions.

While \citet{baumgartner2022} evaluates whether different QCA solution types can reliably recover robustly sufficient conditions, ThS-QCA does not seek to establish robustness in this sense; instead, it offers a systematic and descriptive mapping of how sufficient configurations vary across substantively meaningful outcome and condition thresholds.

\citet{skaaning2011} examined how variations in frequency and consistency thresholds affect QCA results and highlighted the importance of assessing solution stability.

\citet{oana2021} proposed a more comprehensive robustness test protocol, implemented in the \texttt{SetMethods} package. This protocol includes:

\begin{itemize}
\item \textbf{Sensitivity range}: the range over which parameters can be varied without altering the solution structure
\item \textbf{Fit-oriented robustness}: Variation in consistency and coverage
\item \textbf{Case-oriented robustness}: stability of case membership and assignment across solutions
\end{itemize}

However, these existing approaches have a confirmatory character. That is, they evaluate how robust results are to the thresholds set by the researcher.

In contrast, ThS-QCA adopts an exploratory approach. It treats thresholds as ``manipulable variables'' rather than ``fixed values,'' revealing how causal structures are generated, disappear, branch, or reorganize as thresholds change. Table~\ref{tab:confirmatory-exploratory} illustrates this difference.

Importantly, ThS-QCA is not designed as a robustness testing tool in the confirmatory sense. Rather than evaluating whether a single solution remains invariant within a tolerance range, we adopt an exploratory perspective in which thresholds are treated as manipulable analytical variables. This perspective allows researchers to examine how sufficient condition structures emerge, transform, and disappear across substantively meaningful threshold levels.

\begin{table}[ht]
\centering
\small
\caption{Comparison of confirmatory and exploratory approaches}
\label{tab:confirmatory-exploratory}
\begin{tabular}{p{3.5cm}p{4cm}p{4cm}}
\toprule
Aspect & Confirmatory (Existing) & Exploratory (ThS-QCA) \\
\midrule
Meaning of threshold & Range of plausible calibration choices & Qualitatively different theoretical levels \\
Interpretation of change & Problem (vulnerability) & Analytical opportunity \\
Threshold selection & Single theoretically justified threshold & Compare across multiple thresholds \\
Purpose & Validate initial solution & Discover threshold-dependent causal paths \\
\bottomrule
\end{tabular}
\end{table}

\section{Package design}\label{package-design}

\subsection{Design philosophy}\label{design-philosophy}

ThS-QCA is designed according to four fundamental principles:

\textbf{Principle 1: Complementarity}

ThS-QCA builds upon, rather than replaces, existing QCA packages. It uses the \texttt{truthTable()} and \texttt{minimize()} functions provided by the \texttt{QCA} package \citep{thiem2013, dusa2019} directly, adding a new layer of threshold exploration. This design ensures:

\begin{itemize}
\item Consistency with QCA methodology is ensured
\item Automatic benefits from \texttt{QCA} package updates (algorithm improvements, etc.)
\item Output format is familiar to QCA users
\end{itemize}

\textbf{Principle 2: Transparency}

All intermediate results are made available, enabling verification of the analysis process. Truth tables, solution objects, necessity analysis results at each threshold are saved and accessible.

\textbf{Principle 3: Reproducibility}

Complete reconstruction of analyses is enabled. All setting parameters are saved in result objects, and \texttt{generate\_report()} automatically generates reports that include reproducibility information.

\textbf{Principle 4: Task-specific interfaces}

Although a single general-purpose function could technically handle all sweep types through parameter configuration, ThS-QCA provides dedicated functions for each analytical task: outcome threshold sweeps, single-condition sweeps, multi-condition sweeps, and dual sweeps. This design reduces cognitive load, improves computational efficiency by avoiding unnecessary grid expansion, and ensures that output formats are optimized for each specific use case.

From a user perspective, this design minimizes the cognitive and technical cost of conducting large-scale threshold sensitivity analyses, reducing what would otherwise require dozens of manual re-analyses to a single, reproducible function call.

\subsection{Three types of QCA solutions}\label{three-types-of-qca-solutions}

QCA minimization can produce three types of solutions depending on how logical remainders (unobserved configurations) are handled. Following standard QCA terminology \citep{schneider2012, dusa2019}, ThS-QCA supports all three solution types through the \texttt{include} and \texttt{dir.exp} parameters:

\begin{table}[ht]
\centering
\small
\caption{Three types of QCA solutions and their parameter settings}
\label{tab:solution-types}
\begin{tabular}{p{3cm}p{2cm}p{3cm}p{4.5cm}}
\toprule
Solution Type & \texttt{include} & \texttt{dir.exp} & Description \\
\midrule
Complex (Conservative) & \texttt{""} & \texttt{NULL} & Most conservative; uses only observed configurations \\
Parsimonious & \texttt{"?"} & \texttt{NULL} & Most simplified; uses all logical remainders \\
Intermediate & \texttt{"?"} & \texttt{c(1,1,...)} & Theory-guided; uses remainders consistent with directional expectations \\
\bottomrule
\end{tabular}
\end{table}

As of version 2.0.0, the \texttt{ThSQCA} package uses the same default parameter values as \texttt{QCA::minimize()}: \texttt{include = ""} and \texttt{dir.exp = NULL}. This produces the \textbf{complex (conservative) solution} by default, ensuring consistency with the established QCA package.

Researchers can obtain other solution types by explicitly specifying these parameters:

\begin{verbatim}
# Complex solution (default)
result <- otSweep(dat, outcome, conditions, sweep_range, thrX)

# Parsimonious solution
result <- otSweep(dat, outcome, conditions, sweep_range, thrX,
                  include = "?")

# Intermediate solution
result <- otSweep(dat, outcome, conditions, sweep_range, thrX,
                  include = "?", dir.exp = c(1, 1, 1))
\end{verbatim}

\subsection{Relationship to the QCA package}\label{relationship-to-the-qca-package}

\begin{figure}[H]
\centering
\includegraphics[width=0.95\linewidth]{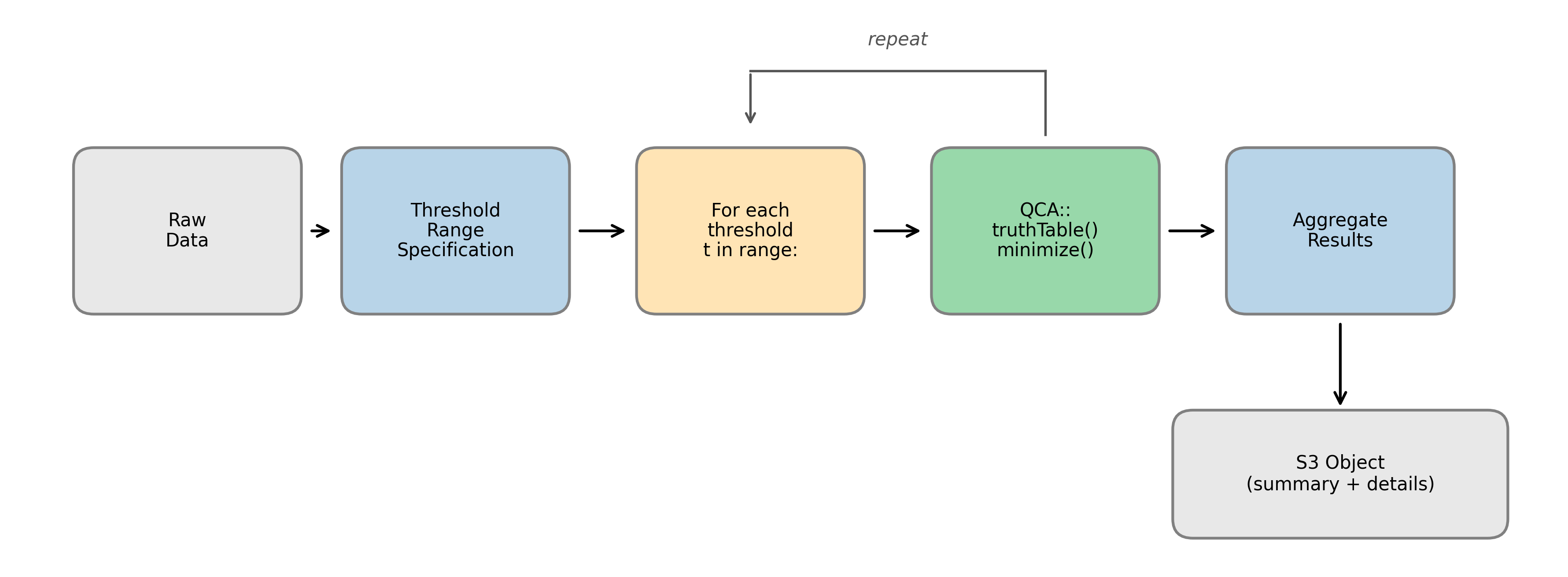}
\caption{ThS-QCA workflow diagram. The package takes raw data and threshold range specifications as input, then iterates through each threshold value. At each iteration, ThS-QCA calls QCA::truthTable() and QCA::minimize() from the QCA package to perform standard QCA analysis. Results are aggregated across all thresholds and returned as an S3 object containing both a summary table for cross-threshold comparison and detailed results for each threshold.}
\label{fig:fig2}
\end{figure}

Figure~\ref{fig:fig2} illustrates the ThS-QCA workflow and its integration with the \texttt{QCA} package. ThS-QCA operates as a wrapper that iteratively calls \texttt{QCA} package functions for each threshold value and aggregates results into a structured S3 object.

Below is simplified R code illustrating this relationship:

\begin{verbatim}
otSweep <- function(dat, outcome, conditions, sweep_range, thrX, 
                    incl.cut = 0.8, n.cut = 1, ...) {
  
  results <- list()
  
  for (t in sweep_range) {
    # Step 1: Dichotomize data at threshold t
    dat_binary <- dichotomize_data(dat, outcome, t, conditions, thrX)
    
    # Step 2: Call QCA package functions
    tt <- QCA::truthTable(
      data = dat_binary,
      outcome = outcome,
      conditions = conditions,
      incl.cut = incl.cut,
      n.cut = n.cut,
      ...
    )
    
    sol <- QCA::minimize(tt, ...)
    
    # Step 3: Extract and save results
    results[[as.character(t)]] <- list(
      threshold = t,
      truth_table = tt,
      solution = sol,
      metrics = extract_metrics(sol)
    )
  }
  
  # Build summary table
  summary_table <- build_summary(results)
  
  structure(
    list(
      summary = summary_table,
      details = results,
      settings = list(sweep_range = sweep_range, thrX = thrX, ...)
    ),
    class = "otSweep"
  )
}
\end{verbatim}

Through this design, ThS-QCA functions as a meta-analytic framework that repeatedly calls QCA.

\subsection{Package structure}\label{package-structure}

The \texttt{ThSQCA} package has the following structure:

\begin{verbatim}
ThSQCA/
|-- R/
|   |-- ThSQCA-package.R      # Package-level documentation
|   |-- tsqca_core.R         # Core utilities (binarization, validation)
|   |-- tsqca_ots_dts.R      # otSweep() and dtSweep()
|   |-- tsqca_cts.R          # ctSweepS() and ctSweepM()
|   |-- tsqca_report.R       # Report generation
|   |-- tsqca_config_chart.R # Configuration chart generation
|   |-- tsqca_helpers.R      # Helper functions
|   |-- tsqca_methods.R      # S3 print/summary methods
|   `-- sample_data.R        # Dataset documentation
|-- data/
|   `-- sample_data.rda      # Sample data
|-- man/                     # Roxygen2-generated documentation
|-- vignettes/
|   |-- ThSQCA_Tutorial_EN.Rmd
|   `-- ThSQCA_Reproducible_EN.Rmd
|-- tests/
|   `-- testthat/            # Unit tests
|-- DESCRIPTION
|-- NAMESPACE
`-- README.md
\end{verbatim}

\subsection{S3 methods}\label{s3-methods}

The use of S3 classes allows users to treat sweep results as analyzable objects, enabling downstream comparison, visualization, and reporting beyond printed solutions.

The \texttt{ThSQCA} package implements S3 methods for all result objects:

\begin{verbatim}
### print method: Display summary table
print.otSweep <- function(x, ...) {
  cat("Outcome Threshold Sweep Results\n")
  cat("================================\n")
  cat("Sweep range:", min(x$settings$sweep_range), "-", 
      max(x$settings$sweep_range), "\n")
  cat("Conditions:", paste(x$settings$conditions, collapse = ", "), "\n")
  cat("Condition thresholds:", 
      paste(names(x$settings$thrX), "=", x$settings$thrX, 
            collapse = ", "), "\n")
  cat("Consistency cutoff:", x$settings$incl.cut, "\n\n")
  print(x$summary)
  invisible(x)
}

### summary method: Detailed statistical information
summary.otSweep <- function(object, ...) {
  cat("Summary of Outcome Threshold Sweep\n")
  cat("===================================\n\n")
  
  # Solution stability analysis
  n_thresholds <- nrow(object$summary)
  unique_solutions <- length(unique(object$summary$expression))
  
  cat("Number of thresholds analyzed:", n_thresholds, "\n")
  cat("Number of unique solutions:", unique_solutions, "\n")
  cat("Solution stability:", 
      round(1 - (unique_solutions - 1) / (n_thresholds - 1), 3), "\n\n")
  
  # Consistency and coverage ranges
  cat("Consistency range:", 
      min(object$summary$inclS), "-", max(object$summary$inclS), "\n")
  cat("Coverage range:", 
      min(object$summary$covS), "-", max(object$summary$covS), "\n")
  
  invisible(object)
}
\end{verbatim}

\subsection{Function overview}\label{function-overview}

Table~\ref{tab:function-overview} summarizes the main functions provided by ThS-QCA.

\begin{table}[ht]
\centering
\small
\caption{ThS-QCA function overview}
\label{tab:function-overview}
\begin{tabular}{p{3.2cm}p{3.8cm}p{5.5cm}}
\toprule
Function & Purpose & Key Arguments \\
\midrule
\texttt{otSweep()} & Outcome threshold sweep & \texttt{sweep\_range}, \texttt{thrX}, \texttt{pre\_calibrated}, \texttt{include}, \texttt{dir.exp} \\
\texttt{ctSweepS()} & Single condition sweep & \texttt{sweep\_var}, \texttt{sweep\_range}, \texttt{thrY}, \texttt{pre\_calibrated}, \texttt{dir.exp} \\
\texttt{ctSweepM()} & Multi-condition sweep & \texttt{sweep\_list}, \texttt{thrY}, \texttt{pre\_calibrated}, \texttt{include}, \texttt{dir.exp} \\
\texttt{dtSweep()} & Dual threshold sweep & \texttt{sweep\_range\_Y}, \texttt{sweep\_list\_X}, \texttt{pre\_calibrated}, \texttt{dir.exp} \\
\texttt{generate\_report()} & Report generation & \texttt{result}, \texttt{output\_file} \\
\bottomrule
\end{tabular}
\end{table}

\subsection{Dependencies and imports}\label{dependencies-and-imports}

The \texttt{ThSQCA} package has minimal dependencies to ensure stability:

\begin{verbatim}
Imports:
    QCA
Suggests:
    testthat (>= 3.0.0),
    knitr,
    rmarkdown
\end{verbatim}

The package imports only the \texttt{QCA} package as a hard dependency, ensuring that users can install and use ThS-QCA with minimal additional packages.

\section{Main functions}\label{main-functions}

This section introduces the main functions provided by the \textbf{ThSQCA} package. All examples use the dataset bundled with the package (\texttt{sample\_data}) and are designed as minimal, fully reproducible demonstrations. Default consistency and frequency cutoffs are used unless explicitly specified. Detailed outputs are computed internally but omitted from display for brevity.

Note: The examples in this section compute the \textbf{intermediate solution} by explicitly specifying \texttt{include = "?"} and \texttt{dir.exp = c(1, 1, 1)}. To obtain the complex solution (default in v2.0.0), omit these parameters.

\begin{verbatim}
library(ThSQCA)
data("sample_data", package = "ThSQCA")
\end{verbatim}

\subsection{Outcome Threshold Sweep: otSweep()}\label{outcome-threshold-sweep-otsweep}

\subsubsection{Purpose}\label{purpose}

The \texttt{otSweep()} function explores how sufficient condition structures change as the outcome threshold varies, while condition thresholds are held constant. This function typically serves as the starting point for threshold-sweep analysis.

\subsubsection{Function signature}\label{function-signature}

\begin{verbatim}
otSweep(
  dat,
  outcome,
  conditions,
  sweep_range,
  thrX,
  pre_calibrated = NULL,
  dir.exp = NULL,
  include = "",
  incl.cut = 0.8,
  n.cut = 1,
  pri.cut = 0,
  extract_mode = c("first", "all", "essential"),
  return_details = TRUE
)
\end{verbatim}

\subsubsection{Example with bundled sample data}\label{example-with-bundled-sample-data}

\begin{verbatim}
res_ots <- otSweep(
  dat            = sample_data,
  outcome        = "Y",
  conditions     = c("X1", "X2", "X3"),
  sweep_range    = 6:8,
  thrX           = c(X1 = 7, X2 = 7, X3 = 7),
  include        = "?",
  dir.exp        = c(1, 1, 1),
  return_details = TRUE
)

summary(res_ots)
#> OTS-QCA Summary
#> =============== 
#> 
#> Analysis Parameters:
#>   Outcome: Y 
#>   Conditions: X1, X2, X3 
#>   Consistency cutoff: 0.8 
#>   Frequency cutoff: 1 
#> 
#> Results by Threshold:
#> 
#>  thrY  expression inclS  covS n_solutions
#>     6  X3 + X1*X2 0.906 0.853           1
#>     7  X3 + X1*X2 0.906 0.879           1
#>     8 No solution    NA    NA           0

# Detailed results are computed internally
# but omitted from the output for brevity.
\end{verbatim}

\subsubsection{Reading the output}\label{reading-otsweep-output}

The summary table contains one row per threshold value tested. The columns are interpreted as follows:

\begin{itemize}
\item \textbf{thrY}: The outcome threshold value applied at that iteration.
\item \textbf{expression}: The minimized solution formula. The operator \texttt{*} denotes logical conjunction (AND), \texttt{+} denotes logical disjunction (OR), and a leading \texttt{\textasciitilde} denotes negation (absence of the condition). \texttt{No solution} indicates that no configuration met the consistency and frequency cutoffs at that threshold.
\item \textbf{inclS}: Solution consistency. Values of 0.8 or above are conventionally required for sufficiency claims \citep{schneider2012}.
\item \textbf{covS}: Solution coverage. Higher values indicate greater empirical relevance of the solution.
\item \textbf{n\_solutions}: The number of logically equivalent solutions returned by Boolean minimization. When this exceeds 1, the \texttt{extract\_mode} argument controls how they are handled (see Section~\ref{handling-multiple-solutions}).
\end{itemize}

Three outcome patterns are commonly encountered:

\textbf{Stable solution}: The same expression appears across multiple threshold values, indicating that the finding is robust to threshold variation. In the example above, \texttt{X3 + X1*X2} is returned at both \texttt{thrY = 6} and \texttt{thrY = 7}.

\textbf{Changing solution}: Different expressions appear at different thresholds. This signals threshold sensitivity and warrants further investigation using \texttt{ctSweepS()} or \texttt{ctSweepM()} to identify the source of instability.

\textbf{No solution}: No configuration satisfies the cutoffs at the given threshold. This typically occurs when the threshold is set too high, leaving too few cases in the outcome set to support consistent configurations.

\subsection{Single-Condition Threshold Sweep: ctSweepS()}\label{single-condition-threshold-sweep-ctsweeps}

\subsubsection{Purpose}\label{purpose-1}

The \texttt{ctSweepS()} function examines sensitivity to the threshold of a single condition, while holding all other condition thresholds and the outcome threshold fixed.

\subsubsection{Function signature}\label{function-signature-1}

\begin{verbatim}
ctSweepS(
  dat,
  outcome,
  conditions,
  sweep_var,
  sweep_range,
  thrY,
  thrX_default = 7,
  pre_calibrated = NULL,
  dir.exp = NULL,
  include = "",
  incl.cut = 0.8,
  n.cut = 1,
  pri.cut = 0,
  extract_mode = c("first", "all", "essential"),
  return_details = TRUE
)
\end{verbatim}

\subsubsection{Example}\label{example}

\begin{verbatim}
res_cts <- ctSweepS(
  dat            = sample_data,
  outcome        = "Y",
  conditions     = c("X1", "X2", "X3"),
  sweep_var      = "X3",
  sweep_range    = 6:9,
  thrY           = 7,
  thrX_default   = 7,
  include        = "?",
  dir.exp        = c(1, 1, 1),
  return_details = TRUE
)

summary(res_cts)
#> CTS-QCA Summary
#> =============== 
#> 
#> Analysis Parameters:
#>   Outcome: Y 
#>   Conditions: X1, X2, X3 
#>   Consistency cutoff: 0.8 
#>   Frequency cutoff: 1 
#> 
#> Results by Threshold:
#> 
#>  threshold expression inclS  covS n_solutions
#>          6      X1*X2 1.000 0.303           1
#>          7 X3 + X1*X2 0.906 0.879           1
#>          8 X3 + X1*X2 1.000 0.818           1
#>          9 X3 + X1*X2 1.000 0.515           1
\end{verbatim}

\subsubsection{Reading the output}\label{reading-ctsweeps-output}

The \texttt{threshold} column records the value applied to \texttt{sweep\_var} at each iteration; all other condition thresholds remain fixed at \texttt{thrX\_default}. The remaining columns (\texttt{expression}, \texttt{inclS}, \texttt{covS}, \texttt{n\_solutions}) follow the same interpretation as described for \texttt{otSweep()} above.

In the example, the solution changes at \texttt{threshold = 6}: the term \texttt{X3} disappears from the formula, leaving only \texttt{X1*X2}. This indicates that the threshold choice for \texttt{X3} influences whether \texttt{X3} enters the solution. At thresholds 7--9, the solution stabilizes, suggesting robustness across this range.

\subsection{Multiple Condition Threshold Sweep: ctSweepM()}\label{multiple-condition-threshold-sweep-ctsweepm}

\subsubsection{Purpose}\label{purpose-2}

The \texttt{ctSweepM()} function simultaneously varies thresholds for multiple conditions, enabling exploration of the joint threshold space.

\subsubsection{Function signature}\label{function-signature-2}

\begin{verbatim}
ctSweepM(
  dat,
  outcome,
  conditions,
  sweep_list,
  thrY,
  pre_calibrated = NULL,
  dir.exp = NULL,
  include = "",
  incl.cut = 0.8,
  n.cut = 1,
  pri.cut = 0,
  extract_mode = c("first", "all", "essential"),
  return_details = TRUE
)
\end{verbatim}

\subsubsection{Example}\label{example-1}

\begin{verbatim}
res_mcts <- ctSweepM(
  dat            = sample_data,
  outcome        = "Y",
  conditions     = c("X1", "X2", "X3"),
  sweep_list     = list(
    X1 = 6:7,
    X2 = 6:7,
    X3 = 6:7
  ),
  thrY           = 7,
  include        = "?",
  dir.exp        = c(1, 1, 1),
  return_details = TRUE
)

summary(res_mcts)
#> MCTS-QCA Summary
#> ================ 
#> 
#> Analysis Parameters:
#>   Outcome: Y 
#>   Conditions: X1, X2, X3 
#>   Consistency cutoff: 0.8 
#>   Frequency cutoff: 1 
#> 
#> Results by Threshold:
#> 
#>         threshold combo_id     expression inclS  covS n_solutions
#>  X1=6, X2=6, X3=6        1          X1*X2 0.833 0.455           1
#>  X1=7, X2=6, X3=6        2          X1*X2 1.000 0.394           1
#>  X1=6, X2=7, X3=6        3      X1*X2*~X3 0.818 0.273           1
#>  X1=7, X2=7, X3=6        4          X1*X2 1.000 0.303           1
#>  X1=6, X2=6, X3=7        5             X3 0.864 0.576           1
#>  X1=7, X2=6, X3=7        6 ~X1*X3 + X1*X2 0.931 0.818           1
#>  X1=6, X2=7, X3=7        7             X3 0.864 0.576           1
#>  X1=7, X2=7, X3=7        8     X3 + X1*X2 0.906 0.879           1
\end{verbatim}

\subsubsection{Reading the output}\label{reading-ctsweepm-output}

Each row corresponds to one threshold combination, identified by the \texttt{threshold} column (showing all condition threshold values) and a numeric \texttt{combo\_id}. The solution columns are interpreted as before.

Scanning across rows, the analyst can identify which threshold combinations yield the same solution and which produce divergent results. In the example, combinations where \texttt{X3 = 6} (combo\_ids 1--4) tend to produce simpler solutions without \texttt{X3}, while combinations where \texttt{X3 = 7} (combo\_ids 5--8) more consistently include \texttt{X3} in the solution. This pattern suggests that the \texttt{X3} threshold is the primary driver of solution variation across the grid.

\subsection{Dual Threshold Sweep: dtSweep()}\label{dual-threshold-sweep-dtsweep}

\subsubsection{Purpose}\label{purpose-3}

The \texttt{dtSweep()} function jointly varies outcome thresholds and condition thresholds, enabling two-dimensional exploration of the threshold space.

As the most general sweep function, \texttt{dtSweep()} can technically replicate the behavior of the other sweep functions through appropriate parameter settings. However, as discussed in Principle 4, we provide dedicated functions for simpler tasks in order to reduce cognitive load and improve computational efficiency.

\subsubsection{Function signature}\label{function-signature-3}

\begin{verbatim}
dtSweep(
  dat,
  outcome,
  conditions,
  sweep_list_X,
  sweep_range_Y,
  pre_calibrated = NULL,
  dir.exp = NULL,
  include = "",
  incl.cut = 0.8,
  n.cut = 1,
  pri.cut = 0,
  extract_mode = c("first", "all", "essential"),
  return_details = TRUE
)
\end{verbatim}

\subsubsection{Example}\label{example-2}

\begin{verbatim}
res_dts <- dtSweep(
  dat            = sample_data,
  outcome        = "Y",
  conditions     = c("X1", "X2", "X3"),
  sweep_list_X   = list(
    X1 = 6:7,
    X2 = 6:7,
    X3 = 6:7
  ),
  sweep_range_Y  = 6:7,
  include        = "?",
  dir.exp        = c(1, 1, 1),
  return_details = TRUE
)

summary(res_dts)
#> DTS-QCA Summary
#> =============== 
#> 
#> Analysis Parameters:
#>   Outcome: Y 
#>   Conditions: X1, X2, X3 
#>   Consistency cutoff: 0.8 
#>   Frequency cutoff: 1 
#> 
#> Results by Threshold:
#> 
#>  thrY combo_id             thrX     expression inclS  covS n_solutions
#>     6        1 X1=6, X2=6, X3=6          X1*X2 0.833 0.441           1
#>     7        1 X1=6, X2=6, X3=6          X1*X2 0.833 0.455           1
#>     6        2 X1=7, X2=6, X3=6          X1*X2 1.000 0.382           1
#>     7        2 X1=7, X2=6, X3=6          X1*X2 1.000 0.394           1
#>     6        3 X1=6, X2=7, X3=6      X1*X2*~X3 0.818 0.265           1
#>     7        3 X1=6, X2=7, X3=6      X1*X2*~X3 0.818 0.273           1
#>     6        4 X1=7, X2=7, X3=6          X1*X2 1.000 0.294           1
#>     7        4 X1=7, X2=7, X3=6          X1*X2 1.000 0.303           1
#>     6        5 X1=6, X2=6, X3=7             X3 0.864 0.559           1
#>     7        5 X1=6, X2=6, X3=7             X3 0.864 0.576           1
#>     6        6 X1=7, X2=6, X3=7 ~X1*X3 + X1*X2 0.931 0.794           1
#>     7        6 X1=7, X2=6, X3=7 ~X1*X3 + X1*X2 0.931 0.818           1
#>     6        7 X1=6, X2=7, X3=7             X3 0.864 0.559           1
#>     7        7 X1=6, X2=7, X3=7             X3 0.864 0.576           1
#>     6        8 X1=7, X2=7, X3=7     X3 + X1*X2 0.906 0.853           1
#>     7        8 X1=7, X2=7, X3=7     X3 + X1*X2 0.906 0.879           1
\end{verbatim}

\subsubsection{Reading the output}\label{reading-dtsweep-output}

The \texttt{dtSweep()} output extends the \texttt{ctSweepM()} layout with an additional \texttt{thrY} column, creating a two-dimensional view of the threshold space. Each unique \texttt{combo\_id} corresponds to one configuration of condition thresholds; for each \texttt{combo\_id}, multiple rows appear---one per outcome threshold value tested.

Comparing rows with the same \texttt{combo\_id} across different \texttt{thrY} values reveals whether the solution is stable as the outcome threshold changes. Comparing rows with the same \texttt{thrY} across different \texttt{combo\_id} values reveals condition-threshold sensitivity at a fixed outcome level. In the example, the solution \texttt{X3 + X1*X2} at \texttt{combo\_id = 8} is consistent across both \texttt{thrY = 6} and \texttt{thrY = 7}, indicating robustness in two dimensions simultaneously.

\subsection{Notes on output structure and reproducibility}\label{notes-on-output-structure-and-reproducibility}

All sweep functions return structured S3 objects containing:

\begin{itemize}
\item Summary tables for cross-threshold comparison
\item Complete parameter settings
\item Detailed QCA objects (when \texttt{return\_details = TRUE})
\end{itemize}

In this paper, we show only summary outputs to maintain clarity and conciseness. Full details are nonetheless computed internally and remain accessible for reproducibility and further inspection.

\subsection{Mixed crisp/fuzzy analysis: the \texttt{pre\_calibrated} argument (v1.3.0)}\label{pre-calibrated}

Version 1.3.0 introduces the \texttt{pre\_calibrated} argument to all four sweep functions. This argument allows researchers to pass fuzzy-set membership scores---values in $[0, 1]$ produced by \texttt{QCA::calibrate()}---directly to \texttt{QCA::truthTable()} without binarization, while still sweeping other conditions on their original raw scale.

\subsubsection{Motivation}\label{pre-calibrated-motivation}

In some research designs, theoretical considerations justify calibrating certain conditions as fuzzy sets (e.g., a ``youth'' set with a calibrated crossover point), while other conditions are more naturally treated on their original Likert or ratio scale within a threshold-sweep framework. Prior to v1.3.0, ThS-QCA required all conditions to be binarized via \texttt{qca\_bin()}, which prevented this mixed workflow. The \texttt{pre\_calibrated} argument removes this restriction without altering the behavior of existing code.

\subsubsection{Usage}\label{pre-calibrated-usage}

Variables listed in \texttt{pre\_calibrated} are passed directly to \texttt{QCA::truthTable()} without binarization. All other conditions are binarized using \texttt{qca\_bin()} at the thresholds specified in \texttt{thrX}, as usual. Pre-calibrated variables do not require a \texttt{thrX} entry, since no binarization is applied to them.

Sweep variables should be kept on their original raw scale. When a variable is swept on its raw scale (e.g., Likert scores 6, 7, 8), each threshold carries direct substantive meaning. Pre-calibrating a sweep variable into fuzzy membership values and then sweeping over those values (e.g., 0.3, 0.5, 0.7) makes interpretation indirect and harder to justify on substantive grounds.

\subsubsection{Validation and warnings}\label{pre-calibrated-validation}

The function raises an error if a variable listed in \texttt{pre\_calibrated} is not found in \texttt{conditions}, or if its values fall outside $[0, 1]$ (which would indicate that a raw-scale variable was passed by mistake). A warning is issued if the variable contains \texttt{NA} values.

If a variable appears in both \texttt{pre\_calibrated} and a sweep list (\texttt{sweep\_list\_X} in \texttt{dtSweep()}, or \texttt{sweep\_list} in \texttt{ctSweepM()}), the function issues a warning and uses the pre-calibrated values, ignoring the sweep thresholds.

\subsubsection{Backward compatibility}\label{pre-calibrated-backward}

The default value \texttt{pre\_calibrated = NULL} preserves v1.2.0 behavior exactly. All existing scripts run without modification.

\section{Demonstration of advanced output and reporting capabilities}\label{demonstration-of-advanced-output-and-reporting-capabilities}

This section demonstrates the advanced output structures and reporting utilities of ThS-QCA.
The purpose is technical: to illustrate how ThS-QCA organizes and reports results from threshold-sweep analyses. All demonstrations use the bundled dataset (\texttt{sample\_data}) to ensure reproducibility.

\begin{verbatim}
library(ThSQCA)
data("sample_data", package = "ThSQCA")
\end{verbatim}

\subsection{Structured result objects}\label{structured-result-objects}

All ThS-QCA sweep functions return structured S3 objects. These objects store (i) analysis parameters (thresholds, cutoffs, and directional expectations) and (ii) cross-threshold summaries of solution structures and fit measures. When \texttt{return\_details = TRUE}, they additionally retain detailed threshold-specific QCA objects (e.g., truth tables and solution objects) for inspection and reporting.

The following example illustrates the structure returned by \texttt{otSweep()}.

\begin{verbatim}
res_ots <- otSweep(
  dat            = sample_data,
  outcome        = "Y",
  conditions     = c("X1", "X2", "X3"),
  sweep_range    = 6:8,
  thrX           = c(X1 = 7, X2 = 7, X3 = 7),
  include        = "?",
  dir.exp        = c(1, 1, 1),
  return_details = TRUE
)

summary(res_ots)
#> OTS-QCA Summary
#> =============== 
#> 
#> Analysis Parameters:
#>   Outcome: Y 
#>   Conditions: X1, X2, X3 
#>   Consistency cutoff: 0.8 
#>   Frequency cutoff: 1 
#> 
#> Results by Threshold:
#> 
#>  thrY  expression inclS  covS n_solutions
#>     6  X3 + X1*X2 0.906 0.853           1
#>     7  X3 + X1*X2 0.906 0.879           1
#>     8 No solution    NA    NA           0
\end{verbatim}

Although only summary information is printed here, the object retains the full set of threshold-specific results when requested (via \texttt{return\_details = TRUE}). These include truth tables, minimized solutions, and associated fit measures.

\subsection{Handling multiple solutions}\label{handling-multiple-solutions}

At a given threshold, Boolean minimization may yield more than one logically equivalent solution. When this occurs, ThS-QCA preserves the multiplicity and internal structure of the solution set rather than collapsing it prematurely.

\begin{verbatim}
# Number of solutions identified at each threshold
res_ots$summary[, c("thrY", "n_solutions")]
#>   thrY n_solutions
#> 1    6           1
#> 2    7           1
#> 3    8           0
\end{verbatim}

In cases with multiple solutions, ThS-QCA distinguishes between EPIs, which are terms shared across all solutions, and SPIs, which vary across solutions. This distinction is maintained within the result object and can be reflected in downstream summaries and configuration tables.

\subsection{Configuration chart generation}\label{configuration-chart-generation}

Beyond single-threshold summaries, ThS-QCA supports cross-threshold comparison of sufficient configurations. The primary representation is a sweep-level configuration chart, inspired by \citet{fiss2011}, that aligns configurations across outcome thresholds: Columns correspond to outcome thresholds and rows correspond to conditions. Standardized symbols are used to encode the role of each condition within the sufficient configurations identified at each threshold: $\bullet$ indicates condition presence, $\otimes$ indicates condition absence, and a blank cell indicates ``don't care''.

This cross-threshold chart provides a compact visualization of how sufficient configurations emerge, persist, or disappear as the outcome threshold changes and thus constitutes a central output of ThS-QCA. Table~\ref{tab:config-chart} illustrates this format using the results from the \texttt{otSweep()} analysis presented earlier.

In the illustrative dataset used here, the identified configurations are relatively simple and involve only positive conditions, yielding a single solution at lower thresholds. This simplicity reflects properties of the data rather than limitations of the ThS-QCA framework. More complex patterns---including negated conditions and multiple equivalent solutions---are fully supported by the underlying output structures and reporting mechanisms.

\begin{table}[ht]
\centering
\small
\caption{Cross-threshold configuration chart of sufficient conditions (solution-term level)}
\label{tab:config-chart}
\begin{tabular}{lcccc}
\toprule
 & thrY = 6 (M1) & thrY = 6 (M2) & thrY = 7 (M1) & thrY = 7 (M2) \\
\midrule
X1 &  & $\bullet$ &  & $\bullet$ \\
X2 &  & $\bullet$ &  & $\bullet$ \\
X3 & $\bullet$ &  & $\bullet$ &  \\
\bottomrule
\multicolumn{5}{l}{\footnotesize $\bullet$ = condition present; $\otimes$ = condition absent; blank = ``don't care''.}
\end{tabular}
\end{table}

The symbol set can be switched to Unicode, ASCII, or LaTeX formats to accommodate different output environments.

Configuration charts are generated at the solution-term level by default, following Fiss (2011) notation, where each column represents one prime implicant. Threshold-level summary charts, which aggregate all configurations at each threshold into a single column, are also available.

\subsection{Automated report generation}\label{automated-report-generation}

The \texttt{generate\_report()} function produces structured Markdown reports directly from ThS-QCA result objects. These reports combine summary tables, configuration charts, and reproducibility information in a single document.

\begin{verbatim}
generate_report(
  result        = res_ots,
  output_file   = "tsqca_demo_report.md",
  title         = "ThSQCA Demonstration Report",
  dat           = sample_data,
  format        = "full",
  include_chart = TRUE,
  chart_symbol_set = "unicode"
)
#> Report generated: tsqca_demo_report.md
\end{verbatim}

The generated report can be used as:

\begin{itemize}
\item A reproducible appendix
\item A starting point for further customization
\end{itemize}

\subsection{Scope and limitations}\label{scope-and-limitations}

This section is intentionally limited to demonstrating output and reporting capabilities. No substantive claims are made about the empirical meaning of the results derived from \texttt{sample\_data}.

The primary objective is to show how ThS-QCA supports transparent, reproducible, and scalable reporting of threshold-sweep analyses within the standard R Markdown workflow.

\section{Comparison with existing packages}\label{comparison-with-existing-packages}

\subsection{Feature comparison}\label{feature-comparison}

Table~\ref{tab:feature-comparison} compares ThS-QCA with existing QCA-related packages.

\begin{table}[ht]
\centering
\small
\caption{Feature comparison ($\bullet$: Native implementation; $\circ$: Available via QCA package; ---: Not available)}
\label{tab:feature-comparison}
\begin{tabular}{p{5cm}ccc}
\toprule
Feature & ThS-QCA & QCA & SetMethods \\
\midrule
\textbf{Core QCA Functions} & & & \\
\quad Truth table construction & $\circ$ & $\bullet$ & $\circ$ \\
\quad Boolean minimization & $\circ$ & $\bullet$ & $\circ$ \\
\quad Necessity analysis & $\circ$ & $\bullet$ & $\bullet$ \\
\midrule
\textbf{Threshold Sweep} & & & \\
\quad Outcome threshold sweep & $\bullet$ & --- & --- \\
\quad Condition threshold sweep & $\bullet$ & --- & --- \\
\quad Dual threshold sweep & $\bullet$ & --- & --- \\
\midrule
\textbf{Output \& Visualization} & & & \\
\quad Configuration table generation & $\bullet$ & --- & --- \\
\quad Markdown/LaTeX export & $\bullet$ & --- & --- \\
\quad EPI/SPI identification & $\bullet$ & --- & --- \\
\bottomrule
\end{tabular}
\end{table}

\subsection{Positioning}\label{positioning}

ThS-QCA is positioned as a complementary extension to the QCA ecosystem:

\begin{itemize}
\item \textbf{QCA package}: Provides core computational infrastructure (truth tables, minimization)
\item \textbf{SetMethods}: Provides advanced QCA procedures (theory evaluation, MMR, visualization)
\item \textbf{ThSQCA}: Provides systematic threshold exploration (sweep analysis, hierarchical sufficiency)
\end{itemize}

The three packages serve different but complementary purposes. Researchers can use \texttt{QCA} for standard analysis, \texttt{SetMethods} for robustness checks based on chosen thresholds, and \texttt{ThSQCA} for exploring how results change across threshold ranges.

ThS-QCA is not intended as a robustness-testing alternative to SetMethods, but as an exploratory tool that is naturally used before or alongside confirmatory robustness checks.

\textbf{Unique output capabilities}: While \texttt{QCA} and \texttt{SetMethods} focus on analytical functions, ThS-QCA additionally provides publication-ready output generation. Configuration tables in Fiss notation ($\bullet$/$\otimes$) can be generated directly in LaTeX format for journal submissions, and comprehensive Markdown reports facilitate transparent documentation of threshold sensitivity analyses---features not available in existing packages.

\subsection{Workflow comparison}\label{workflow-comparison}

\textbf{Traditional workflow (without ThS-QCA):}

\begin{verbatim}
# Manual iteration required
results <- list()
for (thr in 5:9) {
  # Manually dichotomize data
  dat$Y_bin <- ifelse(dat$Y >= thr, 1, 0)
  dat$X1_bin <- ifelse(dat$X1 >= 7, 1, 0)
  # ... repeat for all conditions
  
  tt <- QCA::truthTable(dat, outcome = "Y_bin", ...)
  sol <- QCA::minimize(tt)
  results[[as.character(thr)]] <- sol
  # Manual aggregation required
}
\end{verbatim}

\textbf{ThS-QCA workflow:}

\begin{verbatim}
# Single function call
result <- otSweep(
  dat = dat,
  outcome = "Y",
  conditions = c("X1", "X2", "X3"),
  sweep_range = 5:9,
  thrX = c(X1 = 7, X2 = 7, X3 = 7)
)

# Structured output ready for analysis
\end{verbatim}

\section{Discussion}\label{discussion}

\subsection{Limitations}\label{limitations}

\textbf{Computational complexity}: The \texttt{dtSweep} function explores a potentially large space. With $k$ conditions and $m$ threshold values each, the condition threshold space alone contains $m^k$ combinations. If outcome thresholds are also varied across $m_Y$ values, the total number of combinations becomes $m_Y \times m^k$. For large $k$ or $m$, computation time may become substantial.

\textbf{Interpretation burden}: More results require more interpretation. Researchers must develop frameworks for making sense of solution patterns across thresholds.

\textbf{Data requirements}: Meaningful threshold sweeps require sufficient variation in the data. If most cases cluster around a limited range of values, varying thresholds may not induce substantive changes in set membership or sufficient configurations.

\subsection{Best practices}\label{best-practices}

Based on our experience, we recommend the following practices:

\begin{enumerate}

\item \textbf{Start with \texttt{otSweep}}: Begin by exploring outcome threshold sensitivity before investigating condition thresholds. This establishes a baseline understanding of how the solution changes with the outcome definition and helps identify a stable outcome threshold range to fix for subsequent condition sweeps.

\begin{verbatim}
# Step 1: single-value test to verify the setup
result_test <- otSweep(
  dat = dat, outcome = "Y",
  conditions = c("X1", "X2", "X3"),
  sweep_range = 7,                  # single value first
  thrX = c(X1 = 7, X2 = 7, X3 = 7),
  include = "?", dir.exp = c(1, 1, 1)
)

# Step 2: expand to the substantively meaningful range
result_ots <- otSweep(
  dat = dat, outcome = "Y",
  conditions = c("X1", "X2", "X3"),
  sweep_range = 5:9,
  thrX = c(X1 = 7, X2 = 7, X3 = 7),
  include = "?", dir.exp = c(1, 1, 1)
)
\end{verbatim}

\item \textbf{Use meaningful threshold ranges}: Select threshold ranges that correspond to substantively meaningful categories. For a 10-point Likert scale, the range \texttt{6:8} represents the upper portion of the scale; thresholds below 5 or above 9 are rarely theoretically defensible. Avoid sweeping the entire numeric range of a variable without substantive justification.

\item \textbf{Manage computational complexity}: For \texttt{ctSweepM()} and \texttt{dtSweep()}, the number of analyses grows as $m^k$ (and $m_Y \times m^k$ for \texttt{dtSweep()}), where $m$ is the number of threshold candidates per condition and $k$ is the number of swept conditions. A sweep with three conditions and three values each already yields 27 combinations; five conditions with five values each yields 3{,}125. Start with a narrow range and expand only after confirming the analysis runs correctly.

\begin{verbatim}
# Manageable: 2 x 2 x 2 = 8 combinations
sweep_list_small <- list(X1 = 6:7, X2 = 6:7, X3 = 6:7)

# Use with caution: 5 x 5 x 5 = 125 combinations
sweep_list_large <- list(X1 = 5:9, X2 = 5:9, X3 = 5:9)
\end{verbatim}

\item \textbf{Document threshold rationale}: Explicitly state why specific thresholds were chosen and what they represent substantively. A threshold of 7 on a 10-point satisfaction scale implies that cases scoring 7 or above are treated as ``highly satisfied''; this should be justified on theoretical or measurement grounds rather than chosen to optimize fit.

\item \textbf{Interpret patterns, not just solutions}: Focus on how solution structures evolve across thresholds, identifying recurring patterns such as hierarchical sufficiency (Section~\ref{the-concept-of-hierarchical-sufficiency}). A solution that appears at every threshold is more credible than one that appears only at a single threshold. When solutions change across thresholds, the analyst should examine which conditions enter or exit the formula and whether this change is theoretically interpretable.

\end{enumerate}

\section{Conclusion}\label{conclusion}

\subsection{Contributions}\label{contributions}

We introduce ThS-QCA (Threshold-Sweep QCA), a methodological framework implemented in the \texttt{ThSQCA} R package, that reconceptualizes thresholds in Qualitative Comparative Analysis as explicit objects of systematic exploration rather than fixed calibration choices. By enabling threshold sweep analyses through a unified and reproducible interface, ThS-QCA makes it possible to map how sufficient condition structures vary across substantively meaningful outcome and condition levels.

The \texttt{ThSQCA} package makes three primary contributions:

\begin{enumerate}
\item \textbf{Methodological innovation}: We reconceptualize thresholds from fixed assumptions as objects of systematic study, enabling researchers to examine how causal structures emerge, disappear, and reorganize as thresholds vary.

\item \textbf{Practical tools}: Four sweep functions and automatic report generation reduce tedious manual processes to efficient single-function calls.

\item \textbf{Analytical perspective}: We demonstrate, through fully reproducible examples, how threshold variation can be examined systematically and summarized in a structured manner, without relying on substantive interpretation.
\end{enumerate}

\subsection{Conceptual implications for QCA practice}\label{conceptual-implications-for-qca-practice}

These implications are conceptual rather than formal theoretical propositions.

The hierarchical sufficiency pattern illustrated in this paper highlights a descriptive relationship between outcome stringency and causal complexity. Higher outcome levels typically require more demanding combinations of conditions, whereas broader explanatory coverage is associated with more modest targets. This relationship provides a framework for understanding how causal complexity varies with outcome stringency.

The trade-off between coverage and consistency across thresholds offers researchers explicit guidance for target-setting decisions. Rather than arbitrarily selecting a single threshold, ThS-QCA enables informed choices based on a comprehensive mapping of the threshold-solution space.

\subsection{Future directions}\label{future-directions}

Several avenues for future development merit attention:

\begin{enumerate}
\item \textbf{Computational efficiency}: Reducing computational burden for large threshold spaces through parallel processing and algorithmic optimization.

\item \textbf{Graphical visualization}: Developing visualization functions such as heatmaps and trajectory plots to complement existing output capabilities.

\item \textbf{Fuzzy-set extension}: Adapting the threshold sweep framework for fsQCA, where calibration anchors (full membership, crossover point, full non-membership) can be systematically varied.

\item \textbf{Integration with SetMethods}: Combining ThS-QCA's exploratory threshold analysis with \texttt{SetMethods}' confirmatory robustness protocols to provide comprehensive robustness verification frameworks.
\end{enumerate}

\subsection{Concluding remarks}\label{concluding-remarks}

ThS-QCA represents a first step toward systematizing threshold sensitivity analysis in QCA research. By making such analyses accessible through user-friendly functions and structured outputs, we aim to elevate threshold examination from an occasional robustness check to a core component of QCA practice.

We encourage researchers to incorporate threshold sensitivity analysis as a standard element of QCA reporting. Doing so enhances the transparency, reliability, and credibility of set-theoretic causal inferences in social science research.

\section*{Acknowledgments}
\addcontentsline{toc}{section}{Acknowledgments}

\begin{itemize}
\item This work was supported by JSPS KAKENHI (Grant Number JP20K01998).
\item The author used Claude (Anthropic) for Japanese-to-English translation assistance and Editage for English language editing. The author takes full responsibility for all content.
\end{itemize}

\section{Computational details}\label{computational-details}

All analyses were performed using:

\begin{verbatim}
R version 4.4.2 (2024-10-31 ucrt)
Platform: x86_64-w64-mingw32
Running under: Windows 11 x64

attached base packages:
[1] stats     graphics  grDevices utils     datasets  methods   base     

other attached packages:
[1] QCA_3.23         admisc_0.39      ThSQCA_2.0.0      
[4] rmarkdown_2.30   rjtools_1.0.18.1
\end{verbatim}

\noindent
\textbf{Execution times}: The demonstrations in Section~\ref{main-functions} completed in approximately 8 seconds total on the test system. Execution times scale with the number of threshold combinations: otSweep (3 thresholds, $\sim$0.8s), ctSweepS (4 thresholds, $\sim$1.1s), ctSweepM (8 combinations, $\sim$2s), and dtSweep (16 combinations, $\sim$4s). Times vary by hardware configuration.

\vspace{1em}

\noindent
\textbf{Data and code availability}:

\begin{itemize}
\item CRAN: \url{https://cran.r-project.org/package=ThSQCA}
\item GitHub: \url{https://github.com/im-research-yt/ThSQCA}
\end{itemize}

\section*{Appendix: Function reference}
\addcontentsline{toc}{section}{Appendix: Function reference}

This appendix provides a quick reference summary of all ThS-QCA functions. Table~\ref{tab:function-summary} lists each function with its purpose and key arguments for ease of lookup.

\begin{table}[ht]
\centering
\small
\caption{ThS-QCA function summary}
\label{tab:function-summary}
\begin{tabular}{p{3.2cm}p{3.8cm}p{5.5cm}}
\toprule
Function & Purpose & Key Arguments \\
\midrule
\texttt{otSweep()} & Outcome threshold sweep & dat, outcome, conditions, sweep\_range, thrX, pre\_calibrated, dir.exp \\
\texttt{ctSweepS()} & Single condition sweep & dat, outcome, conditions, thrY, sweep\_var, sweep\_range, pre\_calibrated, dir.exp \\
\texttt{ctSweepM()} & Multi-condition sweep & dat, outcome, conditions, thrY, sweep\_list, pre\_calibrated, dir.exp \\
\texttt{dtSweep()} & Dual threshold sweep & dat, outcome, conditions, sweep\_range\_Y, sweep\_list\_X, pre\_calibrated, dir.exp \\
\texttt{generate\_report()} & Report generation & result, output\_file, title, dat \\
\bottomrule
\end{tabular}
\end{table}

\bibliographystyle{plainnat}
\bibliography{ThSQCA}

\end{document}